\documentclass[useAMS,usenatbib]{mn2e}
\usepackage{newtxtext,newtxmath}
\usepackage{multirow}
\usepackage{times}
\usepackage{graphics,epsfig,txfonts}
\usepackage{textcomp}

\usepackage[T1]{fontenc}
\usepackage{ae,aecompl}
\usepackage[usenames, dvipsnames]{color}
\usepackage[flushleft]{threeparttable}

\usepackage{natbib}

\usepackage{graphicx}	
\usepackage{amsmath}	
\usepackage{amssymb}	
\usepackage{pdflscape}
\usepackage{float}

\title[HMXBs in the LMC]{Identification of High Mass X-ray Binaries selected from XMM-Newton observations of the LMC\thanks{These observations are based on SALT proposal 2013-2-RSA\_OTH\_UKSC-004.}}

\author[N. van Jaarsveld et al.]{
N. van Jaarsveld$^{1,2}$\thanks{E-mail: naomi@saao.ac.za},
D. A. H. Buckley$^{2}$,
V. A. McBride$^{1,2,3}$,
F. Haberl$^{4}$,
\newauthor
G. Vasilopoulos$^{4}$,
C. Maitra$^{4}$,
A. Udalski$^{5}$,
B. Miszalski$^{2,6}$
\\
$^{1}$Department of Astronomy, University of Cape Town, Private Bag X3, Rondebosch, 7701, South Africa\\
$^{2}$South African Astronomical Observatory, PO Box 9, Observatory, 7935, South Africa\\
$^{3}$IAU Office of Astronomy for Development, Cape Town, South Africa\\
$^{4}$Max-Planck-Institut f\"ur extraterrestrische Physik, Giessenbachstra{\ss}e, 85748 Garching, Germany\\
$^{5}$Warsaw University Observatory, Al. Ujazdowskie 4, 00-478 Warszawa, Poland\\
$^{6}$Southern African Large Telescope Foundation, PO Box 9, Observatory, 7935, South Africa}

\date{Accepted 2017 December 14. Received 2017 December 13; in original form 2017 November 3}

\pubyear{2017}

\newcommand{\xmm}{{\it XMM-Newton}}

\begin{document}
\label{firstpage}
\pagerange{\pageref{firstpage}--\pageref{lastpage}}
\maketitle

\begin{abstract}
The Large Magellanic Cloud (LMC) currently hosts around 23 high mass X-ray binaries (HMXBs) of which most are Be/X-ray binaries.  The LMC XMM-Newton survey provided follow-up observations of previously known X-ray sources that were likely HMXBs, as well as identifying new HMXB candidates.  In total 19 candidate HMXBs were selected based on their X-ray hardness ratios. In this paper we present red and blue optical spectroscopy, obtained with SALT and the SAAO 1.9-m telescope, plus a timing analysis of the long term optical light curves from OGLE to confirm the nature of these candidates. We find that 9 of the candidates are new Be/X-ray Binaries, substantially increasing the LMC Be/X-ray binary population. Furthermore, we present the optical properties of these new systems, both individually and as a group of all the BeXBs identified by the XMM-Newton survey of the LMC.
\end{abstract}

\begin{keywords}
 Magellanic Clouds -- X-rays: binaries -- stars: emission-line, Be
\end{keywords}



\section{Introduction}


High Mass X-ray Binaries (HMXBs) include supergiant X-ray binaries (SGXBs) and Be/X-ray binaries (BeXBs).  SGXBs are luminosity class I-II stars, with their 
mass transfer mechanisms being either Roche lobe overflow or wind accretion.  BeXBs making up the largest fraction of HMXBs, having a Oe or Be companion with luminosity class III-V. These stars are either late type O or early B, emission line stars, with the majority having spectral types of B0-B1 (\citealt{Negueruela2002}, \citealt{Coe2005}, \citealt{McBride2008}).  The emission originates from a decretion disc around the star, which formed due to the rapid rotation (rotating at $\gtrsim$75\% \citep{Rivinius2013} of their critical velocity) of the stars, although the mechanism responsible for the decretion disc formation is poorly understood \citep{Porter2003}.

The Magellanic Clouds (MCs) are our closest neighbouring galaxies. As such they provide an independent astrophysical laboratory outside of the Milky Way,  with minimal extinction and known distances.  As such they can be used for extensive studies of their stellar populations and, potentially, the environmental impact on star formation (SF) (\citealt{Dray2006}, \citealt{Walter2015}). We can measure the recent SF rate of the MCs directly by means of their HMXB population, since these massive stars must have formed recently \citep{Grimm2003}.  The Small Magellanic Cloud (SMC) contains a large population of HMXBs. A recent census by \citet{Haberl2016} shows that 120 HMXBs have been identified in the SMC, of which more than 60 show X-ray pulsations indicating the spin period of a neutron star.  It provides a comprehensive sample to conduct population synthesis studies to help understand the recent SF, and how the SMC's environment affected it.  Conversely, only 40 HMXBs have been identified in the Large Magellanic Cloud (LMC), of which only 23 have been confirmed \citep{Antoniou2016,Vasilopoulos2016,Vasilopoulos2017}.  With so few confirmed HMXBs in the LMC it is difficult to make comparisons with population synthesis and SF studies.  More complete samples of LMC HMXBs will allow one to investigate the effects of metallicity on SF at better spatial resolution scales, by comparing the SF history of the SMC (Z$_{SMC}\sim$0.2 Z$_\odot$, \citealt{Luck1998}) with the LMC (Z$_{LMC}\sim$0.5 Z$_\odot$, \citealt{Cole2005}) and the Milky Way.  

Thus in an effort to increase the sample of confirmed LMC HMXBs, we investigate the nature of the optical counterparts of 19 HMXB candidates, some of which were identified by previous surveys \citep{Liu2005}, but with improved positions and hardness ratios from the XMM-Newton survey, as well as new candidates identified by the XMM-Newton survey. We use spectra from the Southern African Large Telescope (SALT) to search for H$\alpha$ emission from the decretion disc, blue spectra from the SAAO 1.9 m telescope to determine the spectral class of the central star, and finally we conduct a timing analysis using $I$-band OGLE IV light curves.  Based on the X-ray and optical properties we confirm 9 of the candidates as BeXBs.  In section 2 we give an overview of the observations, in section 3 we describe the analysis procedures, followed by a discussion of the results in section 4, and a outline of our conclusions in section 5. 

\section{Overview of Observations}

\subsection{XMM data}
\label{sec:data:xmm}

\begin{figure}
  \resizebox{\hsize}{!}{\includegraphics[angle=-90,clip=]{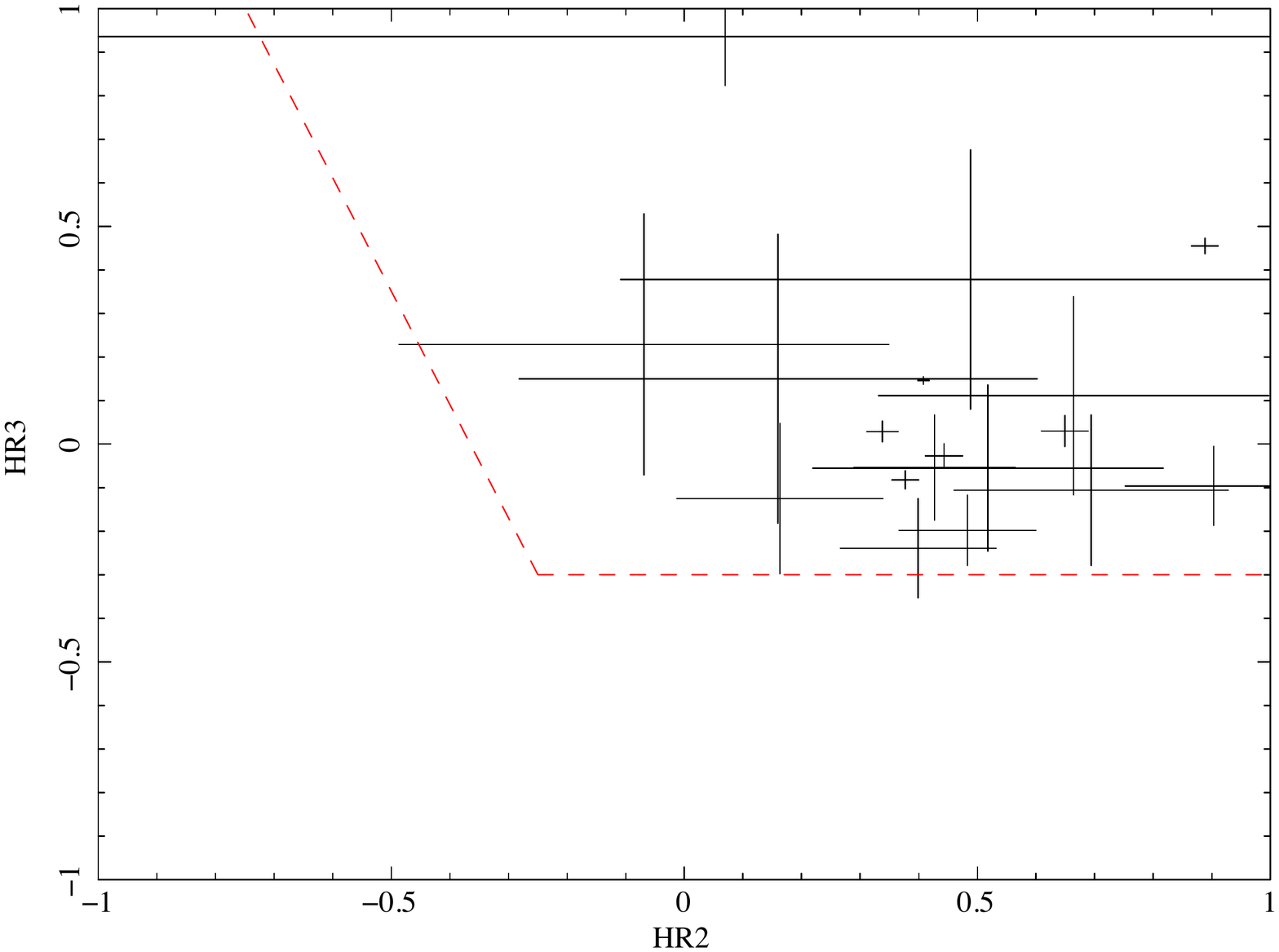}}
  \caption{Hardness ratio diagram for the selected HMXB candidates. The dashed lines define the selection criteria in the 
  HR3 vs. HR2 parameter space applied by \citet{Sturm2013} for Be/X-ray binaries in the Small Magellanic Cloud.}
  \label{fig:xmmhr}
\end{figure}

We selected our targets for optical follow-up spectroscopy from preliminary source detection lists, which we obtained
from \xmm\ \citep{Jansen2001} observations. For our analysis we used the observations from the LMC survey, a 
very large \xmm\ programme to cover the central part of the LMC (PI: F. Haberl), which included archival, as well as ToO observations up to October 2013 within 4\degr\ centred at R.A. = 05h 22m 00s, Dec. = -68\degr\ 30\arcmin\ 00\arcsec\ (see Figure~\ref{fig:xmmhr} in \citealt{Maggi2016}).
\xmm\  carries three X-ray telescopes with a CCD detetcor $-$ the European Photon Imaging Camera (EPIC-pn, EPIC-MOS; \citealt{Struder2001}; \citealt{Turner2001}, respectively) $-$ in each focal plane.
To derive X-ray source detection lists for each observation, we followed the analysis steps of \citet{Sturm2013}, 
 applying a maximum likelihood algorithm (\textsc{edetect\_chain}) from the \xmm\ Science Analysis 
Software\footnote{SAS, http://xmm.esa.int/sas/} to images from the three cameras, simultaneously. 

Candidates for HMXBs are characterised by hard X-ray spectra, which allows them to be selected using hardness ratios (X-ray colours; \citealt{Sturm2013}).  In addition their optical counterparts are bright (typically $V$ magnitudes between 13 and 17 
for stars in the Magellanic Clouds). From our source lists we selected 19 candidates which fulfilled hardness ratio 
criteria (see below) and had a possible optical counterpart within 5\arcsec\ with appropriate optical brightness and colours.
Finally we added Swift J0513.4-6547, a discovered Be/X-ray binary pulsar \citep{Krimm2009} to our sample. 
The target list is presented in Table~\ref{tab:xmmcand} with X-ray properties obtained from re-processing the data with 
the latest version of SAS (16.1) and including new data, which became available since our first detection run. 
To improve the X-ray coordinates, we applied astrometric corrections, using a catalogue of background AGN as the reference frame.
In contrast to \citet{Sturm2013}, we applied background filtering individually for each instrument, which
gains exposure time for EPIC-MOS. 
Candidate 5, originally detected with very low existence likelihood could not be confirmed as an X-ray source and was 
excluded from the list.

We performed source detection simultaneously on images from the three instruments and the five standard energy bands 0.2$-$0.5\,keV, 0.5$-$1\,keV, 
1$-$2\,keV, 2$-$4.5\,keV, and 4.5$-$12\,keV and computed four hardness ratios 
HR$_i$ = (R$_{\rm i+1}$ $-$ R$_{\rm i}$)/(R$_{\rm i + 1}$ + R$_{\rm i}$),
with R$_{\rm i}$ the inferred source counts in band i. Several of our candidates are detected multiple times.
In some cases more than 20 detections are available due to 
frequent calibration observations undertaken of the supernova remnant N132D 
or from a monitoring program of Cal\,83, a supersoft X-ray source in the LMC. 

HR2 and HR3 are listed in Table\,\ref{tab:xmmcand} and plotted in Fig.\,\ref{fig:xmmhr} and were 
derived using the sum of counts from all observations covering the source, i.e. represent average hardness ratios. 
Similarly, average X-ray fluxes are given, which are converted from count rates using conversion factors taken from \citet{Sturm2013}.
Also X-ray coordinates were computed as error-weighted mean and errors (1$\sigma$ confidence, 
with a 0.5\arcsec\ systematic uncertainty added in quadrature) propagated, 
when multiple detections are available. 

A comparison of the X-ray positions with the positions of the selected counterparts from the 
Two Micron All Sky Survey \citep[2MASS; ][]{Skrutskie2006}, shows agreement better than 2$\sigma$ 
for most of the candidates (Table\,\ref{tab:xmmcand}). The largest discrepancy in X-ray to optical 
separation is seen for BeCand-18. This is detected as weak source in 24 observations of N132D, 
in most cases at the rim of the field of view (FoV) with off-axis angles $>$ 13\arcmin, 
which results in relatively large positional uncertainties.  The large number of observations covering BeCand-18 allowed us to reduce the error considerably (Table\,\ref{tab:xmmcand}).
Also, in two observations the source was located at off-axis angles of 6.3\arcmin\ and 8.7\arcmin, 
which also yield separations of 3.9\arcsec\ and 3.1\arcsec, respectively. BeCand-8 and BeCand-9 are also 
located in the FoV of the N132D observations. BeCand-8 is brighter than the other two candidates
and its X-ray and 2MASS position match perfectly, confirming the correct astrometry of these observations.
We conclude, that the large discrepancy in X-ray/optical separation for BeCand-18 makes a chance 
coincidence highly likely.

We took the U, B and V magnitudes of the selected optical counterparts from the Magellanic Clouds Photometric Survey 
\citep[MCPS;][]{Zaritsky2004}, computed U$-$B, B$-$V colours and the reddening-free parameter 
Q = U$-$B $-$ 0.72$\times$(B$-$V) which shows typical values of $-$1.1 to $-$0.7 for the 
known Be/X-ray binaries in the Small Magellanic Cloud \citep[SMC;][]{Haberl2016}. The distribution of 
Q values from our candidate sample (Table\,\ref{tab:xmmcand}) is fully consistent with that of the SMC systems.

\begin{table*}
 \centering
 \begin{minipage}{170mm}
  \caption{HMXBs candidates selected from \xmm\ observations.}
  \label{tab:xmmcand}
   \begin{tabular}{r|rrrrrrr|rr|cr}
   \hline
   \hline
     \multicolumn{1}{c}{Be\footnote{BeCand-5: preliminary X-ray detection not confirmed. 
                                        BeCand-6: no XMM-Newton data, Swift discovery \citep{Krimm2009}. 
                                        BeCand-8/11: coordinates of optical counterpart from 2MASS 6X. 
                                        BeCand-9: No 2MASS catalog entry, coordinates of optical counterpart from OGLE (with Imag). 
                                        BeCand-2: The value for B (16.676 $\pm$ 0.091 mag) is obviously wrong in the MCPS catalogue, therefore we used B = 13.78 mag from Massey (2002) to compute the Q value.  BeCand-18: likely a chance coincidence between optical and X-ray source. }} &
     \multicolumn{7}{c}{----------------------------------------- \xmm\ -----------------------------------------} &
     \multicolumn{2}{c}{---------- 2MASS ----------} &
     \multicolumn{1}{c}{X-2M} &
     \multicolumn{1}{c}{MCPS} \\
     \multicolumn{1}{c}{Cand} &
     \multicolumn{1}{c}{RA} &
     \multicolumn{1}{c}{Dec} &
     \multicolumn{1}{c}{Err} &
     \multicolumn{1}{c}{N} &
     \multicolumn{1}{c}{HR2} &
     \multicolumn{1}{c}{HR3} &
     \multicolumn{1}{c}{Flux} &
     \multicolumn{1}{c}{RA} &
     \multicolumn{1}{c}{Dec} &
     \multicolumn{1}{c}{Sep} &   
     \multicolumn{1}{c}{Q} \\   
     \multicolumn{1}{c}{} &
     \multicolumn{1}{c}{[h m s]} &
     \multicolumn{1}{c}{[\degr\ \arcmin\ \arcsec]} &
     \multicolumn{1}{c}{[\arcsec]} &
     \multicolumn{1}{c}{det} &
     \multicolumn{1}{c}{} &
     \multicolumn{1}{c}{} &
     \multicolumn{1}{c}{[cgs]\footnote{Average (error weighted) 0.2$-$12\,keV flux in erg s$^{-1}$ cm$^{-2}$}} &
     \multicolumn{1}{c}{[h m s]} &
     \multicolumn{1}{c}{[\degr\ \arcmin\ \arcsec]} &
     \multicolumn{1}{c}{[\arcsec]} &
     \multicolumn{1}{c}{[mag]} \\   
   \noalign{\smallskip}\hline\noalign{\smallskip}
     1  & 04 55 46.68 & -69 57 21.0 & 1.47 &   1 & 0.66 $\pm$ 0.33 &  0.11 $\pm$ 0.23 & 1.07E-13 & 04 55 46.34 & -69 57 18.8  &  2.85  &  -1.159 \\   
     2  & 05 00 46.24 & -70 44 35.4 & 0.50 &   1 & 0.41 $\pm$ 0.01 &  0.15 $\pm$ 0.01 & 1.13E-11 & 05 00 46.05 & -70 44 36.0  &  1.14  &  -0.852 \\   
     3  & 05 07 22.33 & -68 47 58.2 & 1.21 &   2 & 0.43 $\pm$ 0.14 & -0.05 $\pm$ 0.12 & 1.01E-13 & 05 07 22.15 & -68 47 59.2  &  1.43  &  -0.515 \\   
     4  & 05 07 55.35 & -68 25 05.1 & 0.27 &   4 & 0.38 $\pm$ 0.02 & -0.08 $\pm$ 0.02 & 1.59E-12 & 05 07 55.47 & -68 25 05.3  &  0.66  &  -1.049 \\   
     6  & 05 13 28.28 & -65 47 18.4 & 0.37 & $-$ &      $-$        &       $-$        &     $-$  & 05 13 28.26 & -65 47 18.7  &  0.35  &  -0.923 \\   
     7  & 05 20 49.11 & -69 19 30.1 & 0.71 &   3 & 0.07 $\pm$ 1.98 &  0.94 $\pm$ 0.11 & 1.50E-13 & 05 20 48.84 & -69 19 30.3  &  1.40  &  -0.576 \\   
     8  & 05 24 17.14 & -69 25 33.7 & 0.24 &  30 & 0.40 $\pm$ 0.13 & -0.24 $\pm$ 0.11 & 8.05E-14 & 05 24 17.15 & -69 25 33.8  &  0.08  &  -0.631 \\   
     9  & 05 25 46.46 & -69 44 50.9 & 0.33 &  22 & 0.48 $\pm$ 0.12 & -0.20 $\pm$ 0.08 & 2.35E-14 & 05 25 46.32 & -69 44 51.7  &  1.11  &  -0.401 \\   
     10 & 05 28 58.38 & -67 09 46.4 & 0.75 &   1 & 0.16 $\pm$ 0.18 & -0.12 $\pm$ 0.17 & 1.18E-13 & 05 28 58.46 & -67 09 45.9  &  0.69  &  -0.871 \\   
     11 & 05 30 10.89 & -69 47 55.8 & 0.91 &   1 & 0.07 $\pm$ 0.42 &  0.23 $\pm$ 0.30 & 2.66E-14 & 05 30 10.79 & -69 47 55.4  &  0.71  &  -0.623 \\   
     12 & 05 30 11.32 & -65 51 23.9 & 0.37 &   2 & 0.34 $\pm$ 0.03 &  0.03 $\pm$ 0.02 & 1.21E-12 & 05 30 11.37 & -65 51 24.1  &  0.34  &  -0.973 \\   
     13 & 05 30 45.42 & -70 40 35.7 & 1.10 &   2 & 0.52 $\pm$ 0.30 & -0.06 $\pm$ 0.19 & 2.35E-14 & 05 30 45.16 & -70 40 34.8  &  1.55  &  -0.573 \\   
     14 & 05 30 59.27 & -68 32 53.4 & 1.68 &   1 & 0.16 $\pm$ 0.44 &  0.15 $\pm$ 0.33 & 1.08E-14 & 05 30 59.41 & -68 32 53.8  &  0.83  &  -0.962 \\   
     15 & 05 31 08.33 & -69 09 23.5 & 0.51 &   1 & 0.89 $\pm$ 0.02 &  0.46 $\pm$ 0.02 & 2.16E-12 & 05 31 08.45 & -69 09 23.5  &  0.64  &  -0.524 \\   
     16 & 05 33 20.87 & -68 41 22.6 & 0.65 &   2 & 0.65 $\pm$ 0.04 &  0.03 $\pm$ 0.04 & 6.47E-13 & 05 33 20.69 & -68 41 23.5  &  1.28  &  -0.945 \\   
     17 & 05 33 28.23 & -67 48 45.9 & 0.82 &   1 & 0.69 $\pm$ 0.23 & -0.11 $\pm$ 0.17 & 1.59E-13 & 05 33 28.19 & -67 48 46.5  &  0.68  &  -0.765 \\   
     18 & 05 25 50.70 & -69 27 29.9 & 0.30 &  24 & 0.90 $\pm$ 0.15 & -0.10 $\pm$ 0.09 & 6.12E-14 & 05 25 50.59 & -69 27 32.9  &  3.02  &  -0.804 \\   
     19 & 05 40 45.48 & -69 14 52.7 & 1.19 &   2 & 0.49 $\pm$ 0.60 &  0.38 $\pm$ 0.30 & 2.07E-14 & 05 40 45.58 & -69 14 51.7  &  1.19  &  -0.772 \\   
     20 & 05 41 34.18 & -68 25 48.5 & 0.12 &  25 & 0.44 $\pm$ 0.03 & -0.03 $\pm$ 0.03 & 2.81E-11 & 05 41 34.32 & -68 25 48.4  &  0.77  &  -0.955 \\   
\hline
\end{tabular}
\end{minipage}
\end{table*}

\subsection{SALT H$\alpha$ spectra}

H$\alpha$ spectra were taken with SALT \citep{Buckley2006} between January 13 2014 and February 8 2014 , and on October 30 2017  with RSS \citep{Burgh2003} in longslit mode, using the PG2300 grating with a slit width of $0.6^{\prime\prime}$, and 2$\times$2 binning, yielding a resolution of $\sim$0.9 \AA\text{ } around H$\alpha$.  The 2-D product spectra received from the SALT pipeline were reduced with PySALT \citep{Crawfod2010}, and wavelength calibrated with a Thorium-Argon lamp. The exposure dates and times for each observation are shown in Table~\ref{tab:instrument}.

\begin{table*}
\centering
\caption{Technical summary of various optical observing campaigns.}
\label{tab:instrument}
\begin{threeparttable}
\begin{tabular}{lccccl}
\hline
\hline
BeCand&\multicolumn{2}{c}{---------------- SALT$^\textit{a}$----------------}&\multicolumn{2}{c}{---------------- 1.9 m$^\textit{b}$ ----------------}& OGLE IV ID\\
&Date&Exp. Time (s)&Date& Exp. Time (s)$^\textit{c}$&\\
\hline
1&13-01-2014&200&24-11-2016&3600&LMC530.26.30737\\
2&13-01-2014&200&24-11-2016&3600&LMC508.31.62\\
3&15-01-2014&400&26-11-2016&5400&LMC510.11.55526\\
4&14-01-2014&200&-&-&LMC510.20.50751\\
6&14-01-2014&250&26-11-2016&3600&LMC506.16.16\\
7&14-01-2014&250&-&-&LMC503.11.56624\\
8&07-02-2014&400&-&-&LMC503.09.85386\\
9&16-01-2014&500&-&-&LMC516.25.32805\\
10&14-01-2014&250&26-11-2016&3600&LMC518.23.4542D\\
11&07-02-2014&500&29-11-2016&5400&-\\
12&15-01-2014&200&24-11-2016&3600&LMC519.29.9750\\
13&08-02-2014&300&27-11-2016&5400&LMC515.31.15984\\
14&30-10-2017&1300&-&-&LMC517.22.4360\\
15&19-01-2014&150&27-11-2016&1800&LMC517.05.18792\\
16&18-01-2014&160&-&-&LMC517.20.21029\\
17&15-01-2014&200&25-11-2016&5400&LMC518.03.9420\\
18&15-01-2014&250&27-11-2016&3600&LMC503.08.66672\\
19&16-01-2014&400&29-11-2016&3600&LMC553.24.87\\
20&16-01-2014&150&27-11-2016&1800&LMC554.06.18693\\
\hline
\end{tabular}
\begin{tablenotes}
\item The $^{\prime\prime}$-$^{\prime\prime}$ signifies the absence of a particular dataset.  $^{\textit{a}}$SALT:  The SALT dataset comprises of all the H$\alpha$ spectra.  $^{\textit{b}}$The 1.9 m dataset comprises of all the blue spectra for spectral classification, $^{\textit{c}}$while the listed exposure times are those of the averaged combined spectra.  
\end{tablenotes}
\end{threeparttable}
\end{table*}

\subsection{SAAO 1.9 m blue spectra}

Blue spectra were taken with the SAAO 1.9\,m telescope between 2016 November 24 and 29.   The observations were carried out with the SpUpNIC spectrograph \citep{Crause2016}, using grating 4 to cover a wavelength range of $\lambda\lambda$3800--5000\,\AA, with 1$\times$2 binning, yielding a resolution of $\sim$4\,\AA.  Depending on the source brightness and the weather conditions, exposure times ranged from 1200 to 1800 seconds, and were observed multiple times if necessary to produce averaged  spectra with SNR of $\sim$40.  Table~\ref{tab:instrument} shows the observation dates and the total exposure times for the average combined spectra.  The spectra were reduced and extracted using IRAF\footnote{http://iraf.noao.edu}, cosmic ray cleaned using a Python based {\tt lacosmic} routine \citep{Dokkum2001}, and wavelength calibrated using Copper-Argon arc lamps.  The final, averaged combined spectra were rectified and smoothed with a 3 point boxcar smoothing routine, using IRAF's {\tt splot} routine. 

\subsection{OGLE IV light curves}

The Optical Gravitational Lensing Experiment \citep{Udalski2015} provided long term I band light curves with roughly daily sampling for 18 of the candidates in the OGLE IV field.  BeCand-11 was the only candidate for which no photometry was available, but the respective OGLE ID's for which photometry was available are listed in Table~\ref{tab:instrument}.  The OGLE light curves were used to study the variability of the candidate optical counterparts.  The identified periodicities and comments on the variability are listed in Table~\ref{tab:halpha}.

\section{Analysis Methods}

\subsection{SALT H$\alpha$ spectra}

\begin{figure}
  \includegraphics[width=\columnwidth]{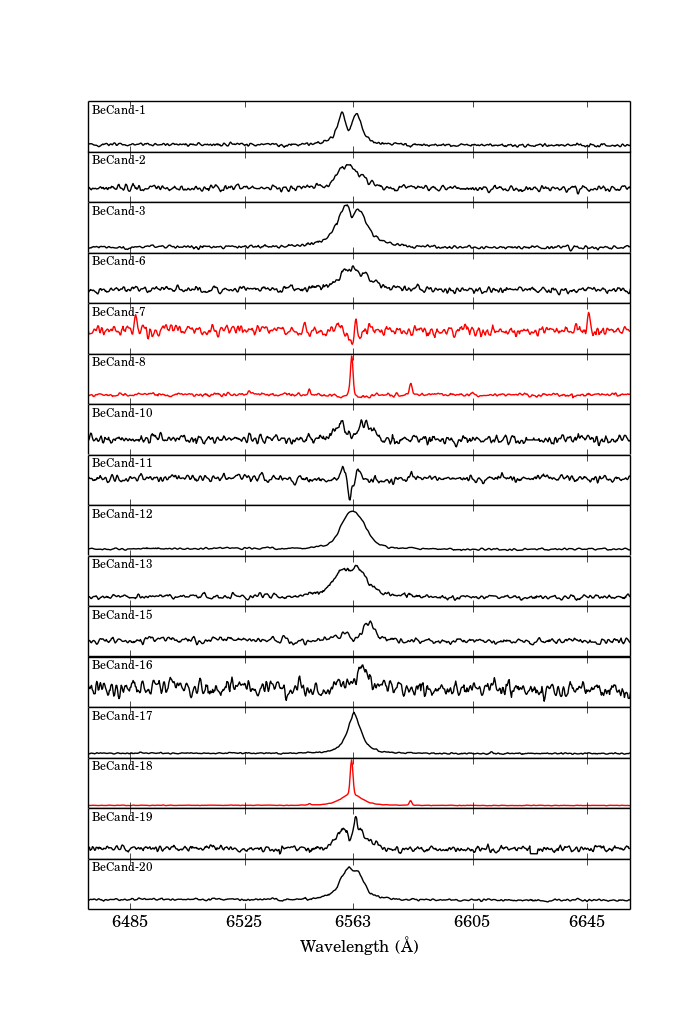}
  \caption{Emission H$\alpha$ profiles.  The y-scale for each profile was chosen independently to highlight the various H$\alpha$ profiles.  The spectra highlighted in red are sources containing narrow H$\alpha$ emission, not necessarily intrinsic to the sources.  BeCand-7 shows H$\alpha$ infilling, however it is not clear if the emission is intrinsic to the source or not.  The H$\alpha$ emission of BeCand-8 is not associated with the optical candidate, but rather with HII region surrounding the optical candidate.  BeCand-18 also highlighted in red shows broad emission from the Be star superimposed on extended narrow emission from a small HII region.  The remaining candidates (black curves) show intrinsic H$\alpha$ emission.}
  \label{fig:paper_halpha}
\end{figure}

Two categories of H$\alpha$ emission were observed namely, narrow emission characteristic of diffuse, interstellar HII regions in the LMC, which is not intrinsic to the candidates, and broad emission resulting from Doppler shifts due to the motion of the decretion disc.  BeCand-8 only showed narrow H$\alpha$ emission, while BeCand-18 had extended narrow H$\alpha$ emission superimposed on broad H$\alpha$ emission (Figure~\ref{fig:paper_halpha}).  Out of the 19 optical candidates 13 candidates exhibited intrinsic H$\alpha$ emission.   IRAF's {\tt splot} was used to measure the equivalent widths (EWs), as well as the V/R ratios for the double peaked profiles.  The H$\alpha$ profile measurements for each candidate was repeated 5 times to determine a average EW and V/R ratio, with the uncertainty estimated from the standard deviation of the mean. Figure~\ref{fig:paper_halpha} shows the emission H$\alpha$ spectra and Table~\ref{tab:halpha} lists the various H$\alpha$ profile measurements for the 15 HMXB candidates that exhibit H$\alpha$ emission.

\begin{center}
\begin{table*}
\centering
\caption{Summary of the results from the optical spectroscopy and OGLE timing study.}
\label{tab:halpha}
\begin{threeparttable}
\begin{tabular}{lcccccccr}
\hline
\hline
BeCand&RA&Dec&V mag&EW(H$\alpha$)&{V/R}&Sp
Type&Period&BeXB?\\
&(J2000)&(J2000)&MCPS&(\AA)&&&(days)\\
\hline
1&04 55 46.34 & -69 57 18.8&14.47&$-22.3 \pm 0.5$&0.997&B1\,IIIe&76 $\pm$ 2&yes\\
2\tnote{\textit{a}}&05 00 46.05 & -70 44 36.0&14.73&$-11 \pm 1$&-&O9\,Ve&30.7 $\pm$ 0.5, 442 $\pm$ 58&yes\\
3&05 07 22.15 & -68 47 59.2&15.79&$-94 \pm 1$&1.230&B3\,IIIe&5.27 $\pm$ 0.02&yes\\
4\tnote{\textit{b}}&05 07 55.35 & -68 25 05.1&14.96&-&-&OB&262 $\pm$ 46&yes\\
6\tnote{\textit{c}}&05 13 28.28 & -65 47 18.4&15.10&$-11.0 \pm 0.3$&-&B0-B1\,Ve&27.4 $\pm$ 0.4&yes\\
7&05 20 49.11 & -69 19 30.1&15.178&$0.7 \pm 0.2$&-&-&440 $\pm$ 43&?\\
8\tnote{\textit{d}}&05 24 17.14 & -69 25 33.7&15.962&$-7 \pm 1$&-&-&32.7 $\pm$ 0.4, 547 $\pm$ 190&?\\
9&05 25 46.46 & -69 44 50.9&16.39&-&-&-&Variability&?\\
10&05 28 58.38 & -67 09 46.4 &15.04&$-8.5 \pm 0.4$&0.952&B0-B0.5\,Ve&193 $\pm$ 48&yes\\
11&05 30 10.89 & -69 47 55.8 &16.07&$-1.2 \pm 0.2$&1.086&B2-B3\,IIIe&-&yes\\
12&05 30 11.32 & -65 51 23.9 &14.88&$-31.2 \pm 0.5$&-&B1-B3\,IIIe-Ve&74 $\pm$ 2&yes\\
13&05 30 45.42 & -70 40 35.7&15.58&$-22.4 \pm 0.4$&-&B0.5-B1\,Ve&280 $\pm$ 44&yes\\
14\tnote{\textit{e}}&05 30 59.27 & -68 32 53.4 &15.91&-&-&-&None&no\\
15\tnote{\textit{f}}&05 31 08.33 & -69 09 23.5&13.70&$-3.5 \pm 0.2$&0.365&B0\,IIIe&Variability&yes\\
16\tnote{\textit{g}}&05 33 20.87 & -68 41 22.6&12.68&-&-&B0.5 Ib&Variability&yes\\
17&05 33 28.23 & -67 48 45.9 &14.82&$-29.4 \pm 0.2$&-&Earlier B0.5\,IIIe&560 $\pm$ 87&yes\\
18\tnote{\textit{h}}&05 25 50.70 & -69 27 29.9 &15.12&$-33.5 \pm 0.9$&-&B2\,IVe-Ve&None&no\\
19&05 40 45.48 & -69 14 52.70&15.98& $-21.6 \pm 0.3$ & 0.740&B1-B3\,IIIe-Ve&Variability&yes\\
20& 05 41 34.18 & -68 25 48.5 &14.04&$-15.5 \pm 0.3$ &-&B0-B1\,IIIe&31.5 $\pm$ 0.4&yes\\
\hline
\end{tabular}
\begin{tablenotes}
\item The coordinates for the optical counterparts are from Table~\ref{tab:xmmcand} listed in the 2MASS column.
\item A $^{\prime\prime}-^{\prime\prime}$ indicate a measurement that could not be made, either due to the absence or the quality of the data.  In the case of the \textit{V/R} column, V/R measurements could only be taken for double peaked H$\alpha$ emission lines.    
\item[\textit{a}] BeCand-2:  Classified as BeXB \citep{Vasilopoulos2016}.
\item[\textit{b}] BeCand-4:  OB spectral class determined from photometry, and classified as BeXB \citep{Maggi2013}.
\item[\textit{c}] BeCand-6:  Classified as BeXB \citep{Coe2015}.
\item[\textit{d}] BeCand-8:  The H$\alpha$ emission does not originate from a decretion disc around the central star, but is from a surrounding nebula.
\item[\textit{e}] BeCand-14:  The H$\alpha$ spectrum was taken in a follow-up campaign with SALT on 30 October 2017.
\item[\textit{f}] BeCand-15: \citet{Vasilopoulos2017} classified the optical companion as a B0 II-Ibe star.
\item[\textit{g}] BeCand-16:  Spectral class determined by \citet{Vasilopoulos2017}.  Possible infilling of the H$\alpha$ line, however the spectrum does not have sufficient SNR to measure a accurate EW.
\item[\textit{h}] BeCand-18: The quoted EW includes only the broad component of the H$\alpha$ emission profile.
\end{tablenotes}
\end{threeparttable}
\end{table*}
\end{center}

\subsection{SAAO 1.9\,m blue spectra}

\begin{center}
\begin{figure*}
\includegraphics[width=\textwidth,height=\textheight,keepaspectratio]{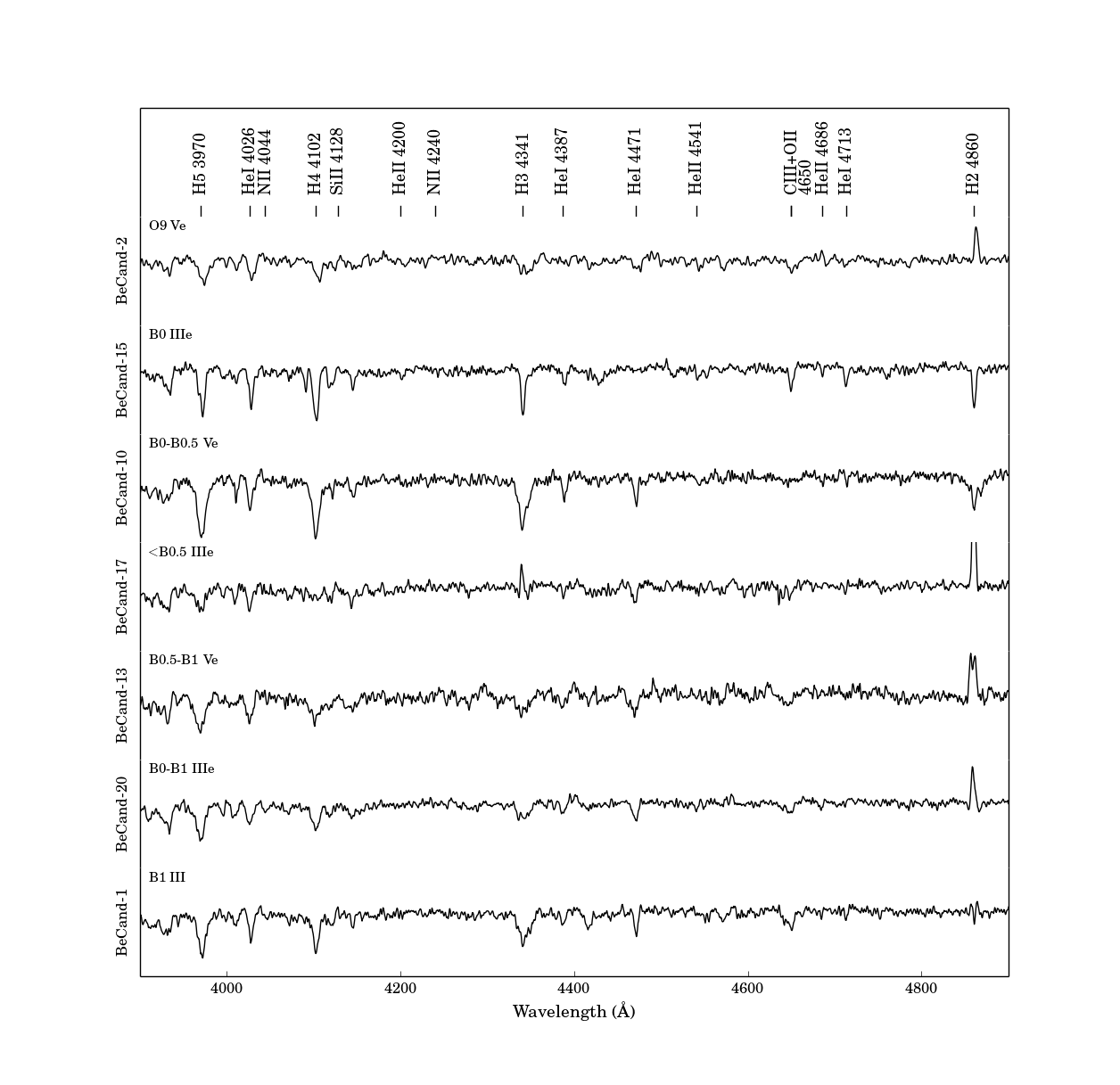}
\caption{Blue-end spectra of candidates with spectral class B1 and earlier.  The open ended emission lines are Balmer emission originating from the decretion disc that were cut off for a better y-scale.}
\label{fig:blue_less_bl}
\end{figure*}
\end{center}

\begin{center}
\begin{figure*}
\includegraphics[width=\textwidth,height=\textheight,keepaspectratio]{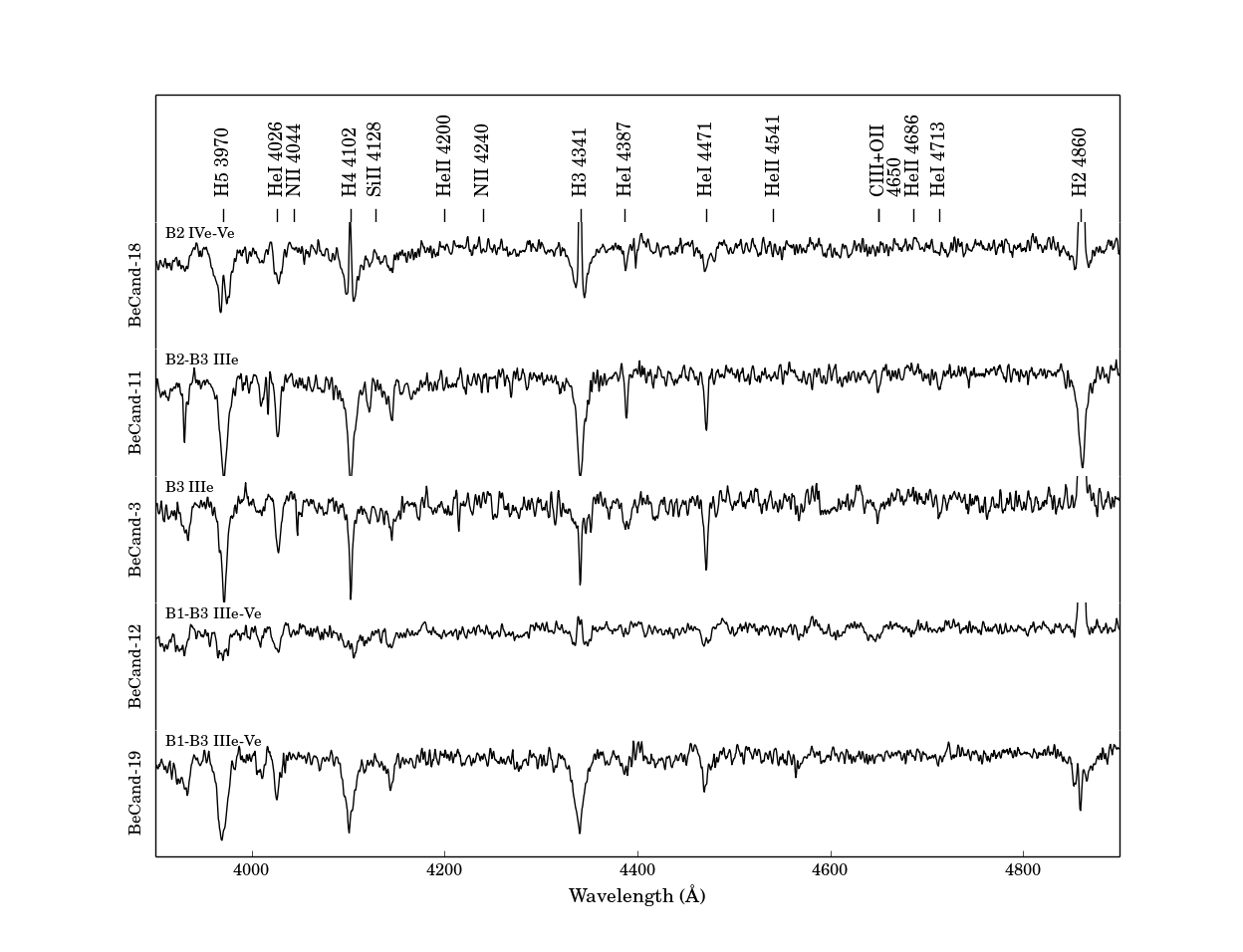}
\caption{Blue spectra of candidates with spectral class later than B2.  The open ended emission lines are Balmer emission originating from the decretion disc that were cut off for a better y-scale.}
\label{fig:blue_b2-b3}
\end{figure*}
\end{center}

The LMC has a lower metallicity than that of the Milky Way (MW), but higher than that of the SMC, hence the spectral behaviour will not necessarily be the same as either the SMC or MW, however \citet{Negueruela2002} and \citet{McBride2008} have shown independently that the spectral type distributions of the LMC and SMC BeXBs are consistent with that of the Milky Way BeXBs.  To determine the spectral class of the candidates we considered spectral classification of massive O and B stars in the SMC, LMC and the MW.  We compared the spectra by eye, using the digital MW spectral atlas provided by \citet{Walborn1990}, the criteria identified by \citet{Evans2004} for the SMC, as well as the spectral atlas of the 30 Doradus region in the LMC by \citet{Walborn2014}, and the criteria provided by \citet{Evans2015} for the north-eastern region of the LMC.  We found that the results were relatively consistent with one another.  The distance modulus resulting from the spectral and luminosity classifications was compared to the distance modulus of the LMC ($\sim$18.48; \citealt{Walker2012}) using the absolute magnitudes from \citet{Zombeck1990}, and direction dependent extinction values from  \citet{Haschke2011}.   Table~\ref{tab:halpha} lists the identified spectral classes, while Figures ~\ref{fig:blue_less_bl} and ~\ref{fig:blue_b2-b3} shows the spectra grouped by spectral class.

\subsection{Timing Analysis}

The timing analysis described here is based on the method applied to SMC BeXBs by \citet{Bird2012}.  The OGLE light curves were detrended with a 51 and 101 day running average routine, acting as a high pass filter to increase the sensitivity of the Lomb-Scargle periodogram at shorter periods.  The Lomb-Scargle periodograms of both the raw and detrended light curves were calculated by $gatspy$ \citep{VanderPlas2015}, a \texttt{Python} based Lomb-Scargle program.
Furthermore, the 3$\sigma$ significance levels of the power spectra were determined by running a Monte Carlo simulation with 10\,000 iterations, using the raw and detrended light curves as inputs.  The simulations were setup such that the time structure was preserved, but the intensities were randomised. Finally, significant aliases, harmonics and beat frequencies were calculated based on the sampling frequency and the annual sampling period to determine which significant peaks in the power spectra were due to variability intrinsic to the source. The timing analysis results are summarised in Table~\ref{tab:halpha}. 

\begin{center}
\begin{figure*}
\includegraphics[width=\textwidth,height=\textheight,keepaspectratio]{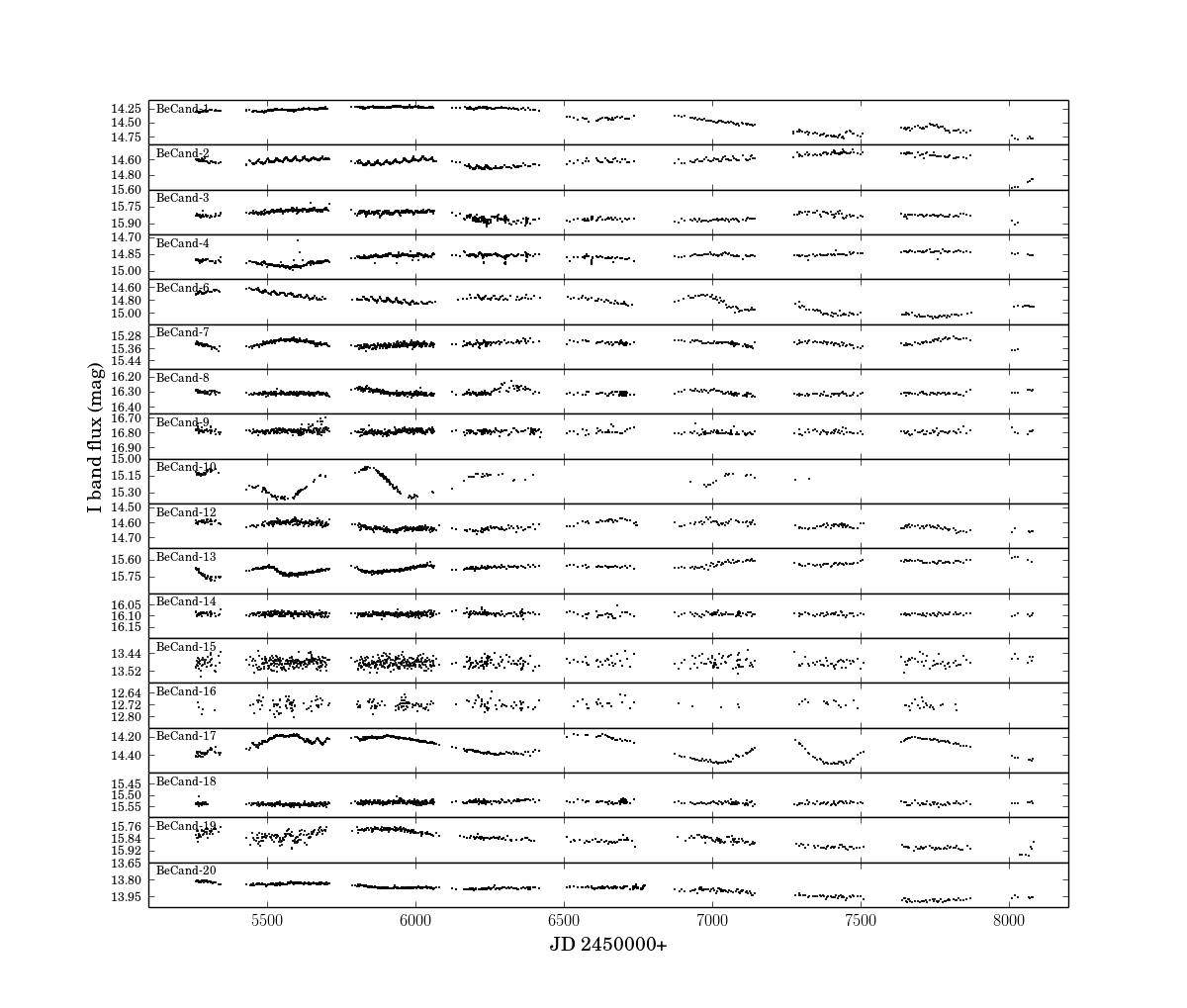}
\caption{OGLE I band light curves.}
\label{fig:ogle}
\end{figure*}
\end{center}
\begin{figure*}
\includegraphics[width=.9\textwidth,height=.9\textheight,keepaspectratio]{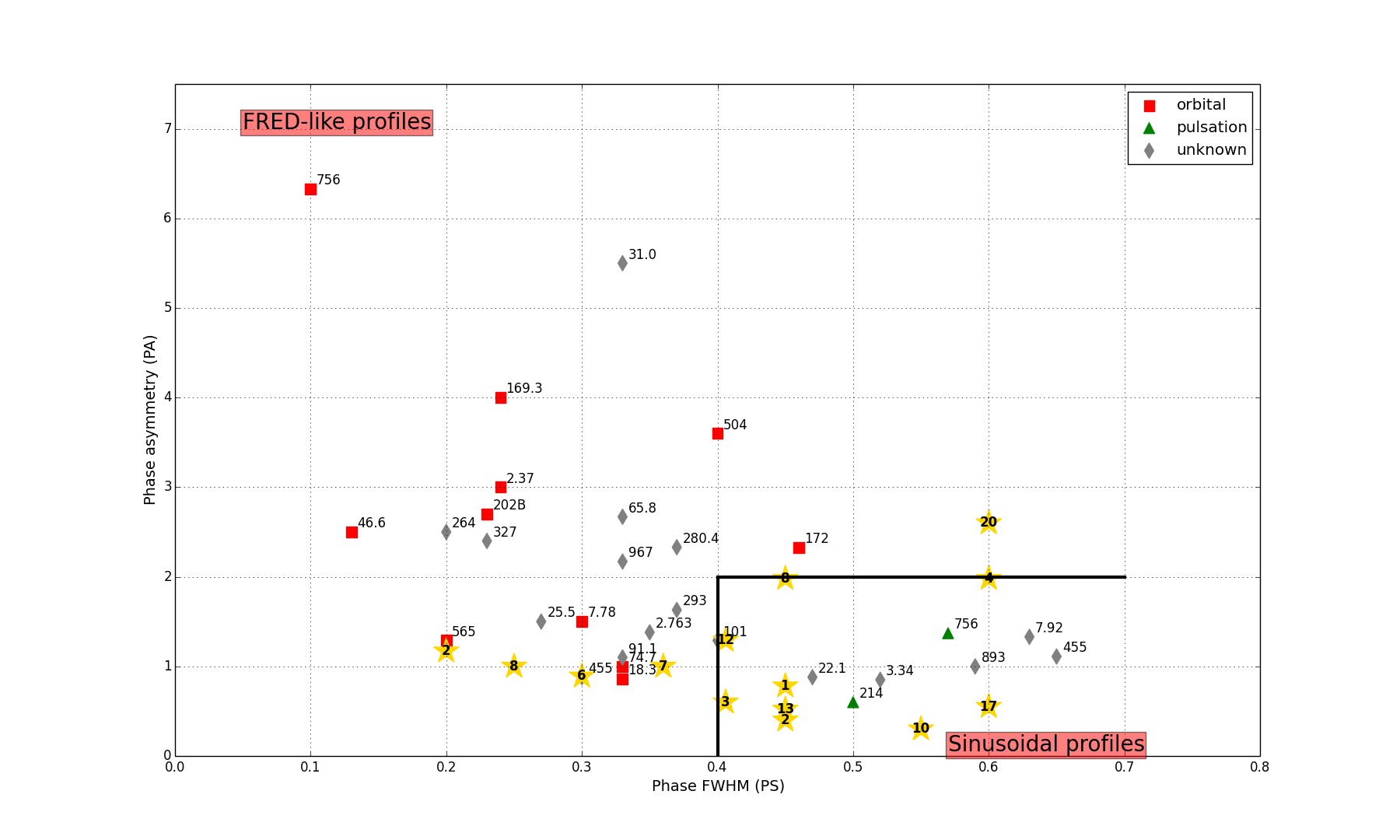}
\caption{Phase metrics to distinguish between orbital pulsations with FRED phased folded light curves, and aliased non-radial pulsations with sinusoidal phased folded light curves.  The number next to each point is the X-ray pulse period in seconds, while our candidates are indicated by the yellow stars with the star number corresponding to the candidate number.  BeCand-2 and 13 in the Sinusoidal region of the plot were shifted slightly in PA in order to resolve the two candidates. Furthermore, BeCand-3,4,8 and 12 are located on the periphery of the FRED and Sinusoidal division.  They were shifted slightly for clarity.   The original plot was taken from \citet{Bird2012}.}
\label{fig:pa-fwhm}
\end{figure*}

\citet{Bird2012} showed that non-radial pulsations (NRPs) of less than a day can beat with the daily sampling period, resulting in significant longer periods (> 10 days) in the power spectra. It is possible to distinguish between aliased NRPs and orbital modulations by considering two metrics from the shape of the phase folded light curves. Since perturbations associated with orbital modulation are caused by the neutron star's interaction with the decretion disc, they result in a fast rise and exponential decay (FRED) phase folded light curve. Aliased NRPs, on the other hand, are associated with a more sinusoidal phase folded light curve.  The metrics are the phase FWHM (PS), and the phase asymmetry (PA).  The PA was given by the fraction of the phase length calculated from the difference between the peak and 10\% of the peak value on the left and right hand side of the phase folded light curve (See Table~\ref{tab:pspa}).  \citet{Bird2012} found that orbital modulation and the aliased NRPs occupy distinct regions in the PA-PS plane (Figure~\ref{fig:pa-fwhm}).  From this we conclude that BeCand-2,6,7,8 and 20 exhibit orbital modulation, while the remaining periodicities appear to be associated with aliased NRPs, except BeCand-1's light curve which showed a clear 76 $\pm$ 2 d modulation with a sinusoidal phase folded light curve.

\begin{table}
\centering
\caption{The phase span (PS) and phase asymmetry (PA) measurements of the phased folded OGLE IV light curves.  The results are plotted in Figure~\ref{fig:pa-fwhm}.}
\label{tab:pspa}
\begin{threeparttable}
\begin{tabular}{ccccc}
\hline
\hline
BeCand &Period& PS (FWHM) & PA    & Type       \\
\hline
1& 76 $\pm$ 2      & 0.45      & 0.78 & Sinusoidal \\
2& 30.7 $\pm$ 0.5      & 0.20      & 1.17 & FRED       \\
-& 442 $\pm$ 58       & 0.45       & 0.40 & Sinusoidal \\
3& 5.27 $\pm$ 0.02      & 0.40      & 0.60 & ? \\
4&262 $\pm$ 46      & 0.60      & 2.00 & ? \\
6&27.4 $\pm$ 0.4      & 0.30       &0.89 & FRED       \\
7&440 $\pm$ 43       & 0.35      & 1.00 & FRED \\
8&32.7 $\pm$ 0.4      & 0.45      & 2.00& ? \\
-&547 $\pm$ 190      & 0.25      & 1.00 & FRED \\
10&193 $\pm$ 48     & 0.55       & 0.30 & Sinusoidal \\
12&74 $\pm$ 2     & 0.40      & 1.29 & ? \\
13&280 $\pm$ 44     & 0.45       & 0.45 & Sinusoidal \\
17&560 $\pm$ 87     & 0.60      & 0.55 & Sinusoidal \\
20&31.5 $\pm$ 0.4     & 0.60      & 2.60 & FRED \\
\hline
\end{tabular}
\begin{tablenotes}
\item The $^{\prime\prime}$?$^{\prime\prime}$ indicates the periods for which the phase folded light curve metrics were located on the periphery of the FRED and Sinusoidal regions of Figure~\ref{fig:pa-fwhm}. 
\end{tablenotes}
\end{threeparttable}
\end{table}

\section{Discussion}

From the \xmm\ survey 19 X-ray sources were identified as potential HMXBs based on their hardness ratio's and the association with an early type star.  In Table~\ref{tab:halpha} we present the spectral properties and timing results of 18 of the optical counterparts that are coincident with the X-ray sources, as well as BeCand-18, which was identified as a likely chance coincidence.

\subsection{New BeXBs}
We classify 9 of the optical candidates as Be stars with spectral types B3 and earlier, with variable and periodic OGLE light curves.  We confirm these 9 sources as new BeXBs, increasing the LMC's BeXB population to 26, moreover increasing the HMXB population of the LMC by $\sim$50\%.   

\subsection{Known XMM BeXBs}

BeCand-2 was detected previously with INTEGRAL (\hbox{IGR~J05007$-$7047}, LXP~38.55) and confirmed as a BeXB with a 38.55\,s X-ray pulsar, and a orbital period of 30.776 $\pm$ 0.005 d \citep{Vasilopoulos2016}, with a B2\,IIIe companion \citep{Masetti2006}.  Our OGLE analysis yielded a consistent orbital period of 30.8 $\pm$ 0.5 d, although the blue optical spectrum had a HeII absorption line at $\lambda\lambda$4200 \AA, which is consistent with a O9e V star.  

Early on in the XXM survey BeCand-4 (LXP 169) was classified as a BeXB based on its X-ray pulse period of 168.8 s, as well as long term X-ray variations, and $I$-band OGLE III-IV light curves.  \citet{Maggi2013} interpreted the rapid decrease in brightness, appearing at all epochs separated by period of 24.328 d, as a binary transit observed at a high inclination angle.  Our Lomb-Scargle analysis revealed a significant period associated with a likely aliased NRP at 262 $\pm$ 46 d (See Figure~\ref{fig:pa-fwhm}).  

Swift observed BeCand-6 (\hbox{Swift J0513.4$-$6547}, LXP 27.2) in April 2009, and was later classified by \citet{Coe2015} as a BeXB with a 27.2\,s pulse period, and a $I$-band OGLE IV orbital period of 27.405 $\pm$ 0.008 d, with $\sim$B1\,Ve companion.  We determined a orbital period of 27.4 $\pm$ 0.4 d from $I$-band OGLE IV light curve, which is consistent with the previous study.  In addition we classified the companion star as a B0e-B1\,Ve star, also consistent with the previous analysis.

\citet{Vasilopoulos2017} confirmed BeCand-15 (XMMU J053108.3-690923) as a HMXB belonging to the SGXB group, with a B0 II-Ibe optical companion and an X-ray spin period of 2013 s. However, in this work we have classified the companion of BeCand-15 as a B0 IIIe star. This discrepancy is likely due to the lower S/N ratio of the blue spectrum which we used for our classification. BeCand-16 (XMMU J053320.8-684122) was also classified as a SGXB with a B0.5 Ibe companion by \citet{Vasilopoulos2017}, with fast flaring X-ray behaviour.

\subsection{Published XMM pulsars}

X-ray pulsations with a period of 272\,s \citep{Haberl2003} was detected for BeCand-12 (\hbox{XMMU~J053011.2$-$655122}).  From our study we confirm BeCand-12 as a BeXB with a significant period around 74 $\pm$ 2 d, with a B1-B3\,III-Ve companion.  

\citet{Shtykovskiy2005} identified BeCand-20 (\hbox{XMMU~J054134.7$-$682550}) as a likely HMXB based on its X-ray properties.  Subsequently \citet{Manousakis2009} observed BeCand-20 in a flaring state and found a 61.601 $\pm$ 0.017 s X-ray pulsation with RXTE.  We confirm BeCand-20 as a BeXB with a 31.5 $\pm$ 0.4 d orbital period, and a B0-B1\,IIIe optical companion.

\subsection{Unconfirmed candidates}
The nature of 3 of the candidates (BeCand-7, 8, 9) remains unresolved, and require additional optical spectroscopic follow up.  BeCand-7 showed definite infilling of the H$\alpha$ line with a EW of 0.7 $\pm$ 0.2, and exhibited an orbital period of 440 $\pm$ 43\,d, however we did not obtain blue-end spectra due to unfavourable weather conditions.  As a result BeCand-7 is likely a Be star based on the data in hand, but the definite spectral type is unknown.  From Figure~\ref{fig:paper_halpha} we observe a narrow H$\alpha$ line for BeCand-8, consistent with diffuse interstellar emission, not necessarily intrinsic to the source, and with a orbital period of 547 $\pm$ 190\,d. The H$\alpha$ spectrum of BeCand-9 was too faint to extract reliably, however the OGLE IV light curve showed some variability.  A second epoch of H$\alpha$ observations will be necessary to determine if these candidates are emission line stars.  In addition to the H$\alpha$ spectra, blue-end spectra are still required to make a reliable spectral classification.  Due to insufficient optical spectroscopy the nature of the optical counterparts remain uncertain for BeCand-7, 8 and 9, however these optical counterparts are coincident with X-ray emission, and in the case for BeCand-7, 8 and 9 were detected multiple times in the XMM data.

\subsection{Non-HMXB systems}
With no H$\alpha$ emission and no variability observed in the OGLE IV light curve, BeCand-14 is likely not at BeXB, while BeCand-18 is peculiar amongst our sample as the H$\alpha$ profile has both broad and narrow components. The narrow component originates from a resolved nebula around 10\arcsec\ in diameter. The emission line ratios of log(H$\alpha$/[N~II])=0.9 and log(H$\alpha$/[S~II])=0.6 are consistent with extragalactic HII regions \citep{Frew2010}. Imaging at mid-infrared wavelengths with \emph{Spitzer} is also consistent with the nebula being an HII region \citep{Meixner2006}. The broad H$\alpha$ component from the B2e IV-V central star suggests a Be star is the ionising source of the HII region. An O-type emission line classification (e.g. \citealt{Walborn2014}) cannot be associated with the broad H$\alpha$ emission given the B2e IV-V classification.

\section{Conclusions}
Eighteen of the XMM X-ray sources were coincident with optical counterparts, of which 3 (BeCand-2, 4 ,6) have been classified as BeXBs, and 2 (BeCand-15, 16) as SGXB  in previous work, while the nature of 3 candidates (BeCand-7,8,9) are still uncertain.  The 9 remaining candidates we classify as new BeXBs (see Table~\ref{tab:halpha}), raising the population of BeXBs in the LMC to 26, and subsequently the number of HMXBs in the LMC to 32.  Moreover, as expected all the BeXBs have spectral classes of B3 and earlier, as well as variable OGLE light curves.

\section*{Acknowledgements}

The XMM-Newton project is supported by the Bundesministerium f\"ur Wirtschaft und 
Technologie/Deutsches Zentrum f\"ur Luft- und Raumfahrt (BMWI/DLR, FKZ 50 OX 0001)
and the Max-Planck Society. 
GV acknowledges support from the BMWI/DLR grant FKZ 50 OR 1208.
DAHB and VM acknowledges support of the National Research Foundation of South Africa (grants IFR2010042800093,98969 and 93405). 
This paper uses observations made at the South African Astronomical Observatory (SAAO) and with the Southern African Large Telescope.  The OGLE project has received funding from the National Science Centre,
Poland, grant MAESTRO 2014/14/A/ST9/00121 to AU.







\bibliographystyle{mn2e}
\bibliography{bexb_paper_2017_new.bib}

\begin{thebibliography}{}

\bibitem[\protect\citeauthoryear{Antoniou \& Zezas}{Antoniou \&
  Zezas}{2016}]{Antoniou2016}
Antoniou V.,  Zezas A.,  2016, MNRAS, 44, 1

\bibitem[\protect\citeauthoryear{Bird, Coe, McBride \& Udalski}{Bird
  et~al.}{2012}]{Bird2012}
Bird A.~J.,  Coe M.~J.,  McBride V.~A.,    Udalski A.,  2012, MNRAS, 423, 3663

\bibitem[\protect\citeauthoryear{Buckley, Swart \& Meiring}{Buckley
  et~al.}{2006}]{Buckley2006}
Buckley D. A.~H.,  Swart G.~P.,    Meiring J.~G.,  2006, in Stepp L.~M.,  ed.,
  Vol. 6267, SPIE. International Society for Optics and Photonics, p. 62670Z

\bibitem[\protect\citeauthoryear{Burgh, Nordsieck, Kobulnicky, Williams,
  O'Donoghue, Smith \& Percival}{Burgh et~al.}{2003}]{Burgh2003}
Burgh E.~B.,  Nordsieck K.~H.,  Kobulnicky H.~A.,  Williams T.~B.,  O'Donoghue
  D.,  Smith M.~P.,    Percival J.~W.,  2003, in Iye M.,  Moorwood A. F.~M.,
  eds,  Vol. 4841, SPIE. International Society for Optics and Photonics,
  p.~1463

\bibitem[\protect\citeauthoryear{Coe, Edge, Galache \& McBride}{Coe
  et~al.}{2005}]{Coe2005}
Coe M.~J.,  Edge W. R.~T.,  Galache J.~L.,    McBride V.~A.,  2005, MNRAS, 356,
  502

\bibitem[\protect\citeauthoryear{Coe, Finger, Bartlett \& Udalski}{Coe
  et~al.}{2015}]{Coe2015}
Coe M.~J.,  Finger M.,  Bartlett E.~S.,    Udalski A.,  2015, MNRAS, 447

\bibitem[\protect\citeauthoryear{Cole, Tolstoy, {Gallagher III} \&
  Smecker-Hane}{Cole et~al.}{2005}]{Cole2005}
Cole A.~A.,  Tolstoy E.,  {Gallagher III} J.~S.,    Smecker-Hane T.~A.,  2005,
  AJ, 129, 1465

\bibitem[\protect\citeauthoryear{Crause et~al.,}{Crause
  et~al.}{2016}]{Crause2016}
Crause L.~A.  et~al., 2016, in Evans C.~J.,  Simard L.,   Takami H.,  eds,
  Vol. 9908, SPIE. p. 990827

\bibitem[\protect\citeauthoryear{Crawford et~al.,}{Crawford
  et~al.}{2010}]{Crawfod2010}
Crawford S.~M.  et~al., 2010. International Society for Optics and Photonics,
  p. 773725

\bibitem[\protect\citeauthoryear{Dray}{Dray}{2006}]{Dray2006}
Dray L.~M.,  2006, MNRAS, 370, 2079

\bibitem[\protect\citeauthoryear{Evans, Howarth, Irwin, Burnley \&
  Harries}{Evans et~al.}{2004}]{Evans2004}
Evans C.~J.,  Howarth I.~D.,  Irwin M.~J.,  Burnley A.~W.,    Harries T.~J.,
  2004, MNRAS, 353, 601

\bibitem[\protect\citeauthoryear{Evans, van Loon, Hainich \& Bailey}{Evans
  et~al.}{2015}]{Evans2015}
Evans C.~J.,  van Loon J.~T.,  Hainich R.,    Bailey M.,  2015, A{\&}A, 584, A5

\bibitem[\protect\citeauthoryear{Frew \& Parker}{Frew \&
  Parker}{2010}]{Frew2010}
Frew D.~J.,  Parker Q.~A.,  2010, PASA, 27, 129

\bibitem[\protect\citeauthoryear{Grimm, Gilfanov \& Sunyaev}{Grimm
  et~al.}{2003}]{Grimm2003}
Grimm H.-J.,  Gilfanov M.,    Sunyaev R.,  2003, MNRAS, 339, 793

\bibitem[\protect\citeauthoryear{Haberl, Dennerl \& Pietsch}{Haberl
  et~al.}{2003}]{Haberl2003}
Haberl F.,  Dennerl K.,    Pietsch W.,  2003, A{\&}A, 406, 471

\bibitem[\protect\citeauthoryear{Haberl \& Sturm}{Haberl \&
  Sturm}{2016}]{Haberl2016}
Haberl F.,  Sturm R.,  2016, A{\&}A, 586

\bibitem[\protect\citeauthoryear{Haschke, Grebel \& Duffau}{Haschke
  et~al.}{2011}]{Haschke2011}
Haschke R.,  Grebel E.~K.,    Duffau S.,  2011, AJ, 141, 158

\bibitem[\protect\citeauthoryear{Jansen et~al.,}{Jansen
  et~al.}{2001}]{Jansen2001}
Jansen F.  et~al., 2001, A{\&}A, 365, 1

\bibitem[\protect\citeauthoryear{Krimm et~al.,}{Krimm et~al.}{2009}]{Krimm2009}
Krimm H.~A.  et~al., 2009, ATel, 2011

\bibitem[\protect\citeauthoryear{Liu, van Paradijs \& van~den Heuvel}{Liu
  et~al.}{2005}]{Liu2005}
Liu Q.~Z.,  van Paradijs J.,    van~den Heuvel E. P.~J.,  2005, A{\&}A, 442,
  1135

\bibitem[\protect\citeauthoryear{Luck, Moffett, {Barnes III} \& Gieren}{Luck
  et~al.}{1998}]{Luck1998}
Luck R.~E.,  Moffett T.~J.,  {Barnes III} T.~G.,    Gieren W.~P.,  1998, AJ,
  115, 605

\bibitem[\protect\citeauthoryear{McBride, Coe, Negueruela, Schurch \&
  McGowan}{McBride et~al.}{2008}]{McBride2008}
McBride V.~A.,  Coe M.~J.,  Negueruela I.,  Schurch M. P.~E.,    McGowan K.~E.,
   2008, MNRAS, 388, 1198

\bibitem[\protect\citeauthoryear{Maggi et~al.,}{Maggi et~al.}{2016}]{Maggi2016}
Maggi P.  et~al., 2016, A{\&}A, 585, A162

\bibitem[\protect\citeauthoryear{Maggi, Haberl, Sturm, Pietsch, Rau, Greiner,
  Udalski \& Sasaki}{Maggi et~al.}{2013}]{Maggi2013}
Maggi P.,  Haberl F.,  Sturm R.,  Pietsch W.,  Rau A.,  Greiner J.,  Udalski
  A.,    Sasaki M.,  2013, A{\&}A, 554

\bibitem[\protect\citeauthoryear{Manousakis, Walter, Audard \& Lanz}{Manousakis
  et~al.}{2009}]{Manousakis2009}
Manousakis A.,  Walter R.,  Audard M.,    Lanz T.,  2009, A{\&}A, 498, 217

\bibitem[\protect\citeauthoryear{Masetti et~al.,}{Masetti
  et~al.}{2006}]{Masetti2006}
Masetti N.  et~al., 2006, A{\&}A, 459, 21

\bibitem[\protect\citeauthoryear{Meixner et~al.,}{Meixner
  et~al.}{2006}]{Meixner2006}
Meixner M.  et~al., 2006, AJ, 132, 2268

\bibitem[\protect\citeauthoryear{Negueruela \& Coe}{Negueruela \&
  Coe}{2002}]{Negueruela2002}
Negueruela I.,  Coe M.~J.,  2002, A{\&}A, 385, 517

\bibitem[\protect\citeauthoryear{Porter \& Rivinius}{Porter \&
  Rivinius}{2003}]{Porter2003}
Porter J.,  Rivinius T.,  2003, PASP, 115, 1153

\bibitem[\protect\citeauthoryear{Rivinius, Carciofi \& Martayan}{Rivinius
  et~al.}{2013}]{Rivinius2013}
Rivinius T.,  Carciofi A.~C.,    Martayan C.,  2013, A{\&}A, 21

\bibitem[\protect\citeauthoryear{Shtykovskiy \& Gilfanov}{Shtykovskiy \&
  Gilfanov}{2005}]{Shtykovskiy2005}
Shtykovskiy P.,  Gilfanov M.,  2005, A{\&}A, 431, 597

\bibitem[\protect\citeauthoryear{Skrutskie et~al.,}{Skrutskie
  et~al.}{2006}]{Skrutskie2006}
Skrutskie M.~F.  et~al., 2006, AJ, 131, 1163

\bibitem[\protect\citeauthoryear{Str{\"{u}}der et~al.,}{Str{\"{u}}der
  et~al.}{2001}]{Struder2001}
Str{\"{u}}der L.  et~al., 2001, A{\&}A, 365, 18

\bibitem[\protect\citeauthoryear{Sturm et~al.,}{Sturm et~al.}{2013}]{Sturm2013}
Sturm R.  et~al., 2013, A{\&}A, 558, A3

\bibitem[\protect\citeauthoryear{Turner et~al.,}{Turner
  et~al.}{2001}]{Turner2001}
Turner M. J.~L.  et~al., 2001, A{\&}A, 365, L27

\bibitem[\protect\citeauthoryear{Udalski, Szyma{\'{n}}ski \&
  Szyma{\'{n}}ski}{Udalski et~al.}{2015}]{Udalski2015}
Udalski A.,  Szyma{\'{n}}ski M.~K.,    Szyma{\'{n}}ski G.,  2015, Acta
  Astronom., 65, 1

\bibitem[\protect\citeauthoryear{van~der Plas, van~den Ancker, Waters \&
  Dominik}{van~der Plas et~al.}{2015}]{VanderPlas2015}
van~der Plas G.,  van~den Ancker M.~E.,  Waters L. B. F.~M.,    Dominik C.,
  2015, A{\&}A, 574, A75

\bibitem[\protect\citeauthoryear{van Dokkum}{van Dokkum}{2001}]{Dokkum2001}
van Dokkum P.~G.,  2001, PASP, 113, 1420

\bibitem[\protect\citeauthoryear{Vasilopoulos, Haberl, Delvaux, Sturm \&
  Udalski}{Vasilopoulos et~al.}{2016}]{Vasilopoulos2016}
Vasilopoulos G.,  Haberl F.,  Delvaux C.,  Sturm R.,    Udalski A.,  2016,
  MNRAS, 461, 1875

\bibitem[\protect\citeauthoryear{Vasilopoulos, Maitra, Haberl, Hatzidimitriou
  \& Petropooulou}{Vasilopoulos et~al.}{2017}]{Vasilopoulos2017}
Vasilopoulos G.,  Maitra C.,  Haberl F.,  Hatzidimitriou D.,    Petropooulou
  M.,  2017, MNRAS

\bibitem[\protect\citeauthoryear{Walborn \& Fitzpatrick}{Walborn \&
  Fitzpatrick}{1990}]{Walborn1990}
Walborn N.~R.,  Fitzpatrick .~L.,  1990, PASP, 102, 379

\bibitem[\protect\citeauthoryear{Walborn et~al.,}{Walborn
  et~al.}{2014}]{Walborn2014}
Walborn N.~R.  et~al., 2014, A{\&}A, 564, A40

\bibitem[\protect\citeauthoryear{Walker}{Walker}{2012}]{Walker2012}
Walker A.~R.,  2012, Ap{\&}SS, 341, 43

\bibitem[\protect\citeauthoryear{Walter, Lutovinov, Bozzo \& Tsygankov}{Walter
  et~al.}{2015}]{Walter2015}
Walter R.,  Lutovinov A.~A.,  Bozzo E.,    Tsygankov S.~S.,  2015, A{\&}A Rev.,
  23, 2

\bibitem[\protect\citeauthoryear{Zaritsky, Harris, Thompson \& Grebel}{Zaritsky
  et~al.}{2004}]{Zaritsky2004}
Zaritsky D.,  Harris J.,  Thompson I.~B.,    Grebel E.~K.,  2004, AJ, 128, 1606

\bibitem[\protect\citeauthoryear{Zombeck}{Zombeck}{1990}]{Zombeck1990}
Zombeck M.,  1990, {Handbook of Space Astronomy and Astrophysics}.
UK: Cambridge University Press

\end{thebibliography}







\bsp	
\label{lastpage}
\end{document}